# Uniqueness and Self-Conjugacy of Dirac Hamiltonians in arbitrary Gravitational Fields

M. V. Gorbatenko, V. P. Neznamov[1]

Russian Federal Nuclear Center – All-Russian Research Institute of Experimental Physics, Sarov, Mira 37, Nizhni Novgorod region, Russia, 607188

Abstract

Proofs of two statements are provided in this paper. First, the authors prove that the formalism of the pseudo-Hermitian quantum mechanics allows describing the Dirac particles motion in arbitrary stationary gravitational fields. Second, it is proved that using the Parker weight operator and the subsequent transition to the $\eta$-representation gives the transformation of the Schrödinger equation for nonstationary metric, when the evolution operator becomes self-conjugate. The scalar products in the $\eta$-representation are flat, which makes possible the use of a standard apparatus for the Hermitian quantum mechanics.

Based on the results of this paper the authors draw a conclusion about solution of the problem of uniqueness and self-conjugacy of Dirac Hamiltonians in arbitrary gravitational fields including those dependent on time.

The general approach is illustrated by the example of Dirac Hamiltonians for several stationary metrics, as well as for the cosmologically flat and the open Friedmann models.

---

[1] E-mail: neznamov@vniief.ru

## 1. INTRODUCTION

In paper [1] we considered the issue of uniqueness and Hermiticity of Hamiltonian for a Dirac particle in weak stationary gravitational field. The relations for Hamiltonians in three systems of tetrad vectors were analyzed using the example of the field described with the Kerr solution: for the system of tetrad vectors used in papers [2]-[4]; for the Killing system of tetrad vectors and for the system of tetrad vectors in the so-called symmetric gauge. It was shown that all the occurring Hamiltonians can be considered using the methods of pseudo-Hermitian quantum mechanics; at that the Hamiltonian in the so-called $\eta$-representation[2] has the form of $H_\eta$, coincident with the Hamiltonian $\tilde{H}_\eta$, occurring during the choice of the system of tetrad vectors, used in papers [2]-[4]. The independence of Hamiltonian $H_\eta$ in the $\eta$-representation on the choice of one of the three systems of tetrad vectors, discovered in [1], does not make it possible to claim that this independence is preserved in the general case as well. Nevertheless, basing on the results of the discussions in [1] we put forward a hypothesis that the Hamiltonian $H_\eta$ in the $\eta$-representation does not depend on the choice of the system of tetrad vectors at all. The additional proof of our supposition was obtained at the analysis of the Parker scalar product [5], [6].

We discovered, that at any choice of the system of tetrad vectors the Hamiltonian $H_\eta$ is expressed via the weight operator

$$\rho = \eta^+ \eta, \qquad (1)$$

used in the Parker scalar product.

During the proving of the Hamiltonian Hermiticity and uniqueness in paper [1] a number of constraints were used. First, the gravitational fields were considered to be weak and stationary. In virtue of this the paper [1] does not suggest a conclusion that the Hamiltonian in the $\eta$-representation is unique in the case of general gravitational fields. Second, we did not observe a connection of the operator $\eta$ with the choice of the system of tetrad vectors used for the description of the Dirac particle dynamics.

In the present paper we eliminate these gaps. Here the issue of uniqueness and self-conjugacy of Dirac Hamiltonians is considered with regard to arbitrary gravitational fields, including those dependent on time.

---

[2] The designations used in [1] are also used in this paper; additional comments on the designations are given in Sections 2, 3.

## 2. FORMALISM OF PSEUDO-HERMITIAN QUANTUM MECHANICS

When we describe the formalism of the pseudo-Hermitian quantum mechanics we follow the works [7]-[9]. The condition of pseudo-Hermiticity of the Hamiltonians assumes the existence of an invertible operator $\rho$ satisfying the relationship

$$\rho H \rho^{-1} = H^+. \tag{2}$$

If there exists an operator $\eta$ satisfying the relationship

$$\rho = \eta^+ \eta, \tag{3}$$

then for the Hamiltonians independent on time we get a Hamiltonian in the $\eta$-representation

$$H_\eta = \eta H \eta^{-1} = H_\eta^+, \tag{4}$$

which is self-conjugate with the spectrum of eigenvalues coincident with the spectrum of the initial Hamiltonian $H$.

The wave function $\psi$ for the initial Hamiltonian satisfies the equation

$$i\frac{\partial \psi}{\partial t} = H\psi, \tag{5}$$

the wave function $\Psi$ in the $\eta$-representation satisfies the equation

$$i\frac{\partial \Psi}{\partial t} = H_\eta \Psi = \eta H \eta^{-1} \Psi, \tag{6}$$

$$\Psi = \eta \psi. \tag{7}$$

The system of units $\hbar = c = 1$ is used in the expressions (5), (6) and in the expressions below.

The scalar product in the initial representation a priori equals to

$$\langle \varphi, \psi \rangle_\rho = \int d^3x \left( \varphi^+ \rho \psi \right). \tag{8}$$

For the wave function in the $\eta$-representation the scalar product has a form standard for the Hermitian quantum mechanics (flat scalar product):

$$(\Phi, \Psi) = \int d^3x \left( \Phi^+ \Psi \right). \tag{9}$$

It is evident that with account for (3), (7) the scalar products of (8) and (9) are:

$$\langle \varphi, \psi \rangle_\rho = (\Phi, \Psi). \tag{10}$$

In paper [1] we also studied the connection of the scalar product of (8) with the Parker scalar product proposed in papers [5], [6]. In the result we discovered that for the three Hamilto-nians of Kerr solutions with different choices of the systems of tetrad vectors, the operator $\rho$ in the expression (8) coincides with the weight operator in the Parker scalar product (1). In Section 4 of this paper this connection is established in the general case for the Dirac Hamiltonian in arbitrary stationary gravitational field with the satisfaction of the pseudo-Hermiticity condition (2).

In the general case of the gravitational fields dependent on time the condition (2) is not satisfied. However, and in this case the transition to the $\eta$-representation is possible with the obtaining of a unique and self-conjugate Hamiltonian with the corresponding flat scalar product. These issues are discussed in Section 6 of this paper.

In Section 8 the algorithm discussed in Sections 6, 7 is used for obtaining self-conjugate Dirac Hamiltonians for several stationary metrics, as well as for nonstationary ones of the cosmologically flat and open Friedmann models.

In the Conclusion the results of this work are discussed.

## 3. REDUCTION OF THE DIRAC EQUATION TO THE FORM OF THE SCHRÖDINGER EQUATION

First, we recall the thread of the corresponding argument and introduce the designations. The tetrad vectors are determined by the relations

$$H_{\underline{\alpha}}^{\mu} H_{\underline{\beta}}^{\nu} g_{\mu\nu} = \eta_{\underline{\alpha}\underline{\beta}}, \tag{11}$$

where

$$\eta_{\underline{\alpha}\underline{\beta}} = \mathrm{diag}[-1,1,1,1]. \tag{12}$$

Three more systems of tetrad vectors $H_{\underline{\alpha}\mu}$, $H^{\underline{\alpha}\mu}$, $H_{\mu}^{\underline{\alpha}}$, can be introduced along with the system of tetrad vectors $H_{\underline{\alpha}}^{\mu}$, which differ from $H_{\underline{\alpha}}^{\mu}$ by the place of the global and local (underlined) indices. Raising and lowering the global indices is performed using the metric tensor $g_{\mu\nu}$ and the inverse tensor $g^{\mu\nu}$; the local indices are raised and lowered using the tensors $\eta_{\underline{\alpha}\underline{\beta}}$, $\eta^{\underline{\alpha}\underline{\beta}}$.

It is assumed that the particle motion is described by the Dirac equation, which in the system of units $\hbar = c = 1$ is written as

$$\gamma^{\alpha} \left( \frac{\partial \psi}{\partial x^{\alpha}} + \Phi_{\alpha} \psi \right) - m\psi = 0. \tag{13}$$

Here $m$ is the particle mass, $\psi$ represents the four-component "column" bispinor, $\gamma^{\alpha}$ are the $4 \times 4$ Dirac matrices, which satisfy the relation

$$\gamma^{\alpha} \gamma^{\beta} + \gamma^{\beta} \gamma^{\alpha} = 2 g^{\alpha\beta} E. \tag{14}$$

In (14) $E$ represents a $4 \times 4$ unity matrix.

Covariant derivative of bispinor $\nabla_{\alpha} \psi$ is in the parenthesis in (13),

$$\nabla_\alpha \psi = \frac{\partial \psi}{\partial x^\alpha} + \Phi_\alpha \psi. \tag{15}$$

The bispinor connectivity $\Phi_\alpha$ is included into the construction (15) for $\nabla_\alpha \psi$; some certain system of tetrad vectors $H_{\underline{\alpha}}^\mu$ determined by the relation (11), should be fixed to retrieve $\Phi_\alpha$. After that the quantity $\Phi_\alpha$ can be expressed via the Christoffel derivatives of the tetrad vectors as follows (the Christoffel derivatives are denoted by semicolon):

$$\Phi_\alpha = -\frac{1}{4} H_\mu^{\underline{\varepsilon}} H_{\underline{\nu\varepsilon};\alpha} S^{\mu\nu}. \tag{16}$$

The expression for $S^{\mu\nu}$ in (16) is determined below – see the formulae (20). The bispinor connectivity $\Phi_\alpha$ in the form of (16) provides the invariance of the covariant derivative $\nabla_\alpha \psi$ with respect to the transition from one system of tetrad vectors to another.

In what follows, we will use Dirac matrices with local indices $\gamma^{\underline{\alpha}}$ along with Dirac matrices with global indices $\gamma^\alpha$. The connection between $\gamma^\alpha$ and $\gamma^{\underline{\alpha}}$ is determined by the relation

$$\gamma^\alpha = H_{\underline{\beta}}^\alpha \gamma^{\underline{\beta}}. \tag{17}$$

It follows from (17), (14), (11) that

$$\gamma^\alpha \gamma^\beta + \gamma^\beta \gamma^\alpha = 2\eta^{\alpha\beta} E. \tag{18}$$

In terms of the matrices $\gamma^{\underline{\alpha}}$ the Dirac equation (13) can be written as follows:

$$H_{\underline{\mu}}^\alpha \gamma^{\underline{\mu}} \left( \frac{\partial \psi}{\partial x^\alpha} + \Phi_\alpha \psi \right) - m\psi = 0. \tag{19}$$

It is convenient to choose the quantities $\gamma^{\underline{\alpha}}$ so that they had the same forms in all local frames of reference. Both the systems $\gamma^{\underline{\alpha}}$ and $\gamma^\alpha$ can be used for building the complete system of $4\times 4$ matrices. An example of a complete system is shown below:

$$E, \quad \gamma_\alpha, \quad S_{\alpha\beta} \equiv \frac{1}{2}(\gamma_\alpha \gamma_\beta - \gamma_\beta \gamma_\alpha), \quad \gamma_5 \equiv \gamma_0 \gamma_1 \gamma_2 \gamma_3, \quad \gamma_5 \gamma_\alpha. \tag{20}$$

Any set of Dirac matrices provides for several discrete automorphisms. We restrict ourselves to the automorphism

$$\gamma_\alpha \to \gamma_\alpha^+ = -D\gamma_\alpha D^{-1}. \tag{21}$$

The matrix $D$ will be called anti-Hermitizing.

The relations (13), (5) suggest that the initial Hamiltonian has the form:

$$H = -\frac{im}{(-g^{00})}\gamma^0 + \frac{i}{(-g^{00})}\gamma^0 \gamma^k \frac{\partial}{\partial x^k} - i\Phi_0 + \frac{i}{(-g^{00})}\gamma^0 \gamma^k \Phi_k. \tag{22}$$

The operator $H$ (22) has the meaning of the evolution operator for the Dirac particle wave function within the chosen global reference frame.

Hereafter we will use the following relations (see, for example, [1]):

$$\gamma_{\underline{\alpha}}^+ = \gamma_{\underline{0}} \gamma_{\underline{\alpha}} \gamma_{\underline{0}}, \quad \gamma_\alpha^+ = \gamma_{\underline{0}} \gamma_\alpha \gamma_{\underline{0}}, \tag{23}$$

$$(\Phi_\alpha)^+ = \gamma_{\underline{0}} \Phi_\alpha \gamma_{\underline{0}}, \tag{24}$$

$$\gamma^0 \gamma^0 = g^{00}, \quad \gamma_{\underline{0}} \gamma_{\underline{0}} = -E. \tag{25}$$

The covariant derivatives of the Dirac matrices are zero:

$$\nabla_\mu \gamma_\alpha = \gamma_{\alpha;\mu} + \left[\Phi_\mu, \gamma_\alpha\right]_- = 0. \tag{26}$$

## 4. Fulfillment of the Condition of Pseudo-Hermiticity with Parker Weight Operator

When the system of tetrad vectors and the external gravitational field are chosen arbitrarily the Dirac Hamiltonian is written in the form of (22). Now let us show, in the general form, in what cases the condition of pseudo-Hermiticity (2) is fulfilled at the use of Parker weight operator

$$\rho = \sqrt{-g}\, \gamma_{\underline{0}} \gamma^0. \tag{27}$$

Direct verification shows that the inverse operator $\rho^{-1}$ has the form

$$\rho^{-1} = \frac{1}{\sqrt{-g}\,(-g^{00})} \gamma^0 \gamma_{\underline{0}}. \tag{28}$$

It is easy to get the evidence that the operator (27) is Hermitian.

$$\rho^+ = \sqrt{-g}\, \gamma^{0+} \gamma_{\underline{0}}^+ = \sqrt{-g}\, \gamma_{\underline{0}} \gamma^0 \gamma_{\underline{0}} \gamma_{\underline{0}} \gamma_{\underline{0}} \gamma_{\underline{0}} = \sqrt{-g}\, \gamma_{\underline{0}} \gamma^0 = \rho. \tag{29}$$

Now let us check the fulfillment of the condition (2) for the Hamiltonian (22) using the operators (27), (28). Let us determine the difference

$$\Delta \equiv H^+ - \rho H \rho^{-1} = \Delta_1 + \Delta_2 + \Delta_3 + \Delta_4, \tag{30}$$

where $\Delta_1, \Delta_2, \Delta_3, \Delta_4$ are defined by the corresponding summands of the Hamiltonian (22).

$$\Delta_1 = \left(-\frac{im}{(-g^{00})} \gamma^0\right)^+ - \sqrt{-g}\, \gamma_{\underline{0}} \gamma^0 \left(-\frac{im}{(-g^{00})}\right) \frac{1}{\sqrt{-g}} \gamma^0 \frac{1}{(-g^{00})} \gamma_{\underline{0}} =$$
$$= im\gamma_{\underline{0}} \gamma^0 \frac{1}{(-g^{00})} \gamma_{\underline{0}} - im\gamma_{\underline{0}} \gamma^0 \frac{1}{(-g^{00})} \gamma_{\underline{0}} = 0; \tag{31}$$

$$\Delta_2 = \left(\frac{i}{(-g^{00})} \gamma^0 \gamma^k \frac{\partial}{\partial x^k}\right)^+ - \sqrt{-g}\, \gamma_{\underline{0}} \gamma^0 \left(\frac{i}{(-g^{00})} \gamma^0 \gamma^k \frac{\partial}{\partial x^k}\right) \frac{1}{\sqrt{-g}} \gamma^0 \frac{1}{(-g^{00})} \gamma_{\underline{0}} =$$
$$= -i\gamma_{\underline{0}} \left(\frac{\partial \gamma^k}{\partial x^k}\right) \gamma^0 \frac{1}{(-g^{00})} \gamma_{\underline{0}} - i\gamma_{\underline{0}} \gamma^k \frac{1}{2(-g)} \frac{\partial(-g)}{\partial x^k} \gamma^0 \frac{1}{(-g^{00})} \gamma_{\underline{0}}; \tag{32}$$

$$\Delta_3 = \left(-i\Phi_0\right)^+ - \sqrt{-g}\,\gamma_{\underline{0}}\gamma^0\left(-i\right)\Phi_0\,\frac{1}{\sqrt{-g}}\,\gamma^0\,\frac{1}{\left(-g^{00}\right)}\gamma_{\underline{0}} =$$
$$= i\gamma_{\underline{0}}\gamma^0_{;0}\gamma^0\,\frac{1}{\left(-g^{00}\right)}\gamma_{\underline{0}};\tag{33}$$

$$\Delta_4 = \left(\frac{i}{\left(-g^{00}\right)}\gamma^0\gamma^k\Phi_k\right)^+ - \sqrt{-g}\,\gamma_{\underline{0}}\gamma^0\left(\frac{i}{\left(-g^{00}\right)}\gamma^0\gamma^k\Phi_k\right)\frac{1}{\sqrt{-g}}\,\gamma^0\,\frac{1}{\left(-g^{00}\right)}\gamma_{\underline{0}} =$$
$$= i\gamma_{\underline{0}}\gamma^k_{;k}\gamma^0\,\frac{1}{\left(-g^{00}\right)}\gamma_{\underline{0}}.\tag{34}$$

Substituting the derived expressions (31)-(34) for $\Delta_1$, $\Delta_2$, $\Delta_3$, $\Delta_4$ in (30) gives:

$$\Delta = H^+ - \rho H \rho^{-1} = i\gamma_{\underline{0}}\begin{pmatrix}\varepsilon\\ \mu\,\varepsilon\end{pmatrix}\gamma^\mu\gamma^0\,\frac{1}{\left(-g^{00}\right)}\gamma_{\underline{0}} + i\gamma_{\underline{0}}\,\frac{\partial\gamma^0}{\partial t}\gamma^0\,\frac{1}{\left(-g^{00}\right)}\gamma_{\underline{0}} -$$
$$-i\gamma_{\underline{0}}\,\frac{1}{2(-g)}\gamma^k\,\frac{\partial(-g)}{\partial x^k}\gamma^0\,\frac{1}{\left(-g^{00}\right)}\gamma_{\underline{0}} =\tag{35}$$
$$= i\gamma_{\underline{0}}\left(\frac{1}{\left(-g^{00}\right)}\,\frac{\partial\gamma^0}{\partial t}\gamma^0 - \frac{1}{2(-g)}\,\frac{\partial(-g)}{\partial t}\right)\gamma_{\underline{0}}.$$

The relation given below was used in (35) for deriving the final expression

$$\begin{pmatrix}\varepsilon\\ \mu\,\varepsilon\end{pmatrix} = \frac{1}{2(-g)}\,\frac{\partial(-g)}{\partial x^\mu}.\tag{36}$$

The expression (35), as it follows from its deriving, is true in the general form, since no any particular suppositions about metric and system of tetrad vectors has been made.

The right-hand part of the relation (35) includes the time derivatives of the metric determinant and of the Dirac matrices $\gamma^0$. Therefore, if these two classes of quantities do not depend on time, then in the case of stationary gravitational fields the pseudo-Hermiticity condition (2) is automatically fulfilled for the Hamiltonian (22). An analogous statement has been proved in [1], but now in this paper this statement is referred to any choice of the system of tetrad vectors and to any stationary gravitational field rather than to three systems of tetrad vectors and to weak Schwarzschild and Kerr solution fields only, which were considered in [1].

In the subsequent discussion we will need to write the relation (35) in the system of tetrad vectors in the Schwinger gauge (see Section 5). At the transition to this system the relation (35) gets changed – the matrix $\gamma^0(x)$ should be replaced by $\tilde{\gamma}^0(x)$ according to the equality $\tilde{\gamma}^0(x) = \sqrt{-g^{00}}\,\gamma^{\underline{0}}$. Thus,

$$\Delta = i\,\frac{\partial\ln\sqrt{-g}}{\partial t} + i\,\frac{\partial\ln\sqrt{-g^{00}}}{\partial t}.\tag{37}$$

Note, that the condition (2) can be fulfilled in some special cases when

$$\frac{1}{(-g^{00})}\frac{\partial \gamma^0}{\partial t}\gamma^0 = \frac{1}{2(-g)}\frac{\partial(-g)}{\partial t}. \tag{38}$$

When the pseudo-Hermiticity condition (2) is fulfilled according to (3)-(10), the transition to the $\eta$ - representation allows obtaining the self-conjugate Hamiltonian (4) with flat scalar product (9) and with the eigenvalue spectrum coincident with the spectrum of initial Hamiltonian $H$.

Now let us show the uniqueness of the Hamiltonian $H_\eta$ determined by the expression (4).

## 5. Uniqueness of the Hamiltonian $H_\eta$

### 5.1. System of Tetrad Vectors in the Schwinger Gauge

In paper [10] Schwinger derived a system of tetrad vectors $\{\tilde{H}^\mu_{\underline{\alpha}}(x)\}$, where the vector $\tilde{H}^0_{\underline{\alpha}}$ contained the following components:

$$\tilde{H}^0_{\underline{0}} \neq 0; \quad \tilde{H}^0_{\underline{k}} = 0. \tag{39}$$

In view of the fact that this system is particularly important, now we are going to systematically describe the procedures and the implications connected with the introduction of (39).

Suppose, that in the considered four-dimensional Riemann space with the signature $(-+++)$ the chosen global reference frame is $\{x^\alpha\}$, and in this system the field of metric tensor $g_{\alpha\beta}(x)$ is specified. Then we suppose that a tangent Minkowski space, a system of tetrad vectors $\{H^\mu_{\underline{\alpha}}(x)\}$ and a system of tetrad Dirac matrices $\{\gamma^{\underline{\alpha}}\}$ constant over the whole space are introduced in each point of the space. The global system of Dirac matrices $\{\gamma^\alpha(x)\}$ is connected with the system $\{\gamma^{\underline{\alpha}}\}$ via the relation (17).

In the three-dimensional sub-space we introduce the tensor

$$f^{mn} \equiv g^{mn} - \frac{g^{0m}g^{0n}}{g^{00}}. \tag{40}$$

Using the equalities followed from the relations $g_{\alpha\varepsilon}g^{\varepsilon\beta} = \delta^\beta_\alpha$ shows that the tensor $f^{mn}$ is an inverse tensor to $g_{mn}$ in the three-dimensional sub-space, i.e. it meets the relations

$$g_{mp}f^{pn} = \delta^n_m, \quad \det(f^{mn}) \neq 0. \tag{41}$$

In the three-dimensional space we introduce an orthonormal system of tetrad vectors $\{\tilde{H}^n_{\underline{m}}\}$, such as those, which satisfy the relations

$$\tilde{H}^m_{\underline{p}}(x)\tilde{H}^n_{\underline{q}}(x)\eta^{\underline{pq}} \equiv \tilde{H}^m_{\underline{p}}(x)\tilde{H}^n_{\underline{p}}(x) = f^{mn}(x). \tag{42}$$

Now we introduce a vector $\tilde{H}^\alpha_{\underline{0}} = (\tilde{H}^0_{\underline{0}}, \tilde{H}^k_{\underline{0}})$ with the components

$$\tilde{H}_{\underline{0}}^0 = \sqrt{-g^{00}}; \quad \tilde{H}_{\underline{0}}^k = -\frac{g^{0k}}{\sqrt{-g^{00}}} \tag{43}$$

and three four-dimensional vectors $\tilde{H}_{\underline{k}}^\mu(x)$ with the components

$$\tilde{H}_{\underline{k}}^0(x) = 0, \quad \tilde{H}_{\underline{k}}^m(x). \tag{44}$$

$\tilde{H}_{\underline{k}}^m(x)$ represents the vectors which satisfy the relations (42). Then the system of four vectors $\tilde{H}_{\underline{\alpha}}^\mu(x) = \{\tilde{H}_{\underline{0}}^\mu(x), \tilde{H}_{\underline{k}}^\mu(x)\}$ is a system of tetrad vectors in the Schwinger gauge.

First we show that the introduced vectors compose a system of tetrad vectors in a four-dimensional space, i.e. meet the constitutive relations

$$\tilde{H}_{\underline{\alpha}}^\mu(x)\tilde{H}_{\underline{\beta}}^\nu(x)\eta^{\underline{\alpha\beta}} = g^{\mu\nu}(x). \tag{45}$$

Now let us write the components of (45) to prove the above said:

$$\left.\begin{array}{l} -\tilde{H}_{\underline{0}}^0\tilde{H}_{\underline{0}}^0 + \tilde{H}_{\underline{k}}^0\tilde{H}_{\underline{k}}^0 = g^{00} \\ -\tilde{H}_{\underline{0}}^0\tilde{H}_{\underline{0}}^m + \tilde{H}_{\underline{k}}^0\tilde{H}_{\underline{k}}^m = g^{0m} \\ -\tilde{H}_{\underline{0}}^m\tilde{H}_{\underline{0}}^n + \tilde{H}_{\underline{k}}^m\tilde{H}_{\underline{k}}^n = g^{mn} \end{array}\right\}. \tag{46}$$

The relations in the first and in the second lines in (46) are obviously satisfied after the values of the components (43), (44) are substituted in them. After (43) is substituted into the third line of (46) we get

$$\tilde{H}_{\underline{k}}^m\tilde{H}_{\underline{k}}^n = g^{mn} - \frac{g^{0m}g^{0n}}{g^{00}}. \tag{47}$$

In the right-hand part of (47) the tensor $f^{mn}$ introduced above can be found. The equalities (47), as well as all the relations (46), are fulfilled because of the vectors $\{\tilde{H}_{\underline{m}}^n\}$ were chosen so that they satisfied the relation (42).

The method of construction makes it clear that the systems of tetrad vectors in the Schwinger gauge are determined with the accuracy up to local spatial rotations in three-dimensional sub-spaces which do not affect the vector $\tilde{H}_{\underline{0}}^\alpha$ with the components (43). At the same time the expression for the vector $\tilde{H}_{\underline{\alpha}}^0$ is unique.

The system of tetrad vectors in the Schwinger representation can be used for building global Dirac matrices. If we designate the Dirac matrices corresponding to $\{\tilde{H}_{\underline{\alpha}}^\mu(x)\}$ via $\{\tilde{\gamma}^\alpha(x)\}$, then according to the general relation (17) we get

$$\tilde{\gamma}^\alpha(x) = \tilde{H}_{\underline{\mu}}^\alpha(x)\gamma^{\underline{\mu}}. \tag{48}$$

Using the relations (43), (44), we get from (48) the following:

$$\left.\begin{array}{l} \tilde{\gamma}^0(x) = \tilde{H}_{\underline{\mu}}^0(x)\gamma^{\underline{\mu}} = \tilde{H}_{\underline{0}}^0(x)\gamma^{\underline{0}} = \sqrt{-g^{00}}\,\gamma^{\underline{0}}, \\ \tilde{\gamma}^k(x) = \tilde{H}_{\underline{0}}^k(x)\gamma^{\underline{0}} + \tilde{H}_{\underline{m}}^k(x)\gamma^{\underline{m}}. \end{array}\right\} \tag{49}$$

The upper line implies that the matrix $\tilde{\gamma}^0(x)$ coincides with $\gamma^{\underline{0}}$ with the accuracy up to the multiplier. This feature distinguishes the system of tetrad vectors in the Schwinger gauge among the other systems.

Note, that when the system of tetrad vectors in the Schwinger gauge is built not all vectors included into the system $\{\tilde{H}^\mu_{\underline{\alpha}}(x)\}$ are defined unambiguously. The relations (43) define the vector $\tilde{H}^\alpha_{\underline{0}} = (\tilde{H}^0_{\underline{0}}, \tilde{H}^k_{\underline{0}})$ unambiguously. As for the vectors $\{\tilde{H}^n_{\underline{m}}\}$, the relations (42) define them with the accuracy up to the spatial rotations in the space with the metric tensor $g_{mn}$ and inverse metric tensor $f^{mn}$. Since these rotations does not affect the vector $\tilde{H}^\alpha_{\underline{0}}$, the generators of spatial rotations commute with $\tilde{\gamma}^0$, and, hence, are the combinations of the matrices $\gamma_{\underline{2}}\gamma_{\underline{3}}, \gamma_{\underline{3}}\gamma_{\underline{1}}, \gamma_{\underline{1}}\gamma_{\underline{2}}$.

## 5.2. Connection between the Arbitrary System of Tetrad Vectors $\{H^\mu_{\underline{\alpha}}(x)\}$ and the System of Tetrad Vectors in the Schwinger Gauge $\{\tilde{H}^\mu_{\underline{\alpha}}(x)\}$

Two arbitrary systems of tetrad vectors in one and the same space are mutually connected by Lorentz transformation. In our case the connection between the systems $\{H^\mu_{\underline{\alpha}}(x)\}$ and $\{\tilde{H}^\mu_{\underline{\alpha}}(x)\}$ is written as

$$\tilde{H}^\mu_{\underline{\alpha}}(x) = \Lambda_{\underline{\alpha}}{}^{\underline{\beta}}(x) H^\mu_{\underline{\beta}}(x). \tag{50}$$

The quantities $\Lambda_{\underline{\alpha}}{}^{\underline{\beta}}(x)$, included into (50), satisfy the relations

$$\left.\begin{array}{l}\Lambda_{\underline{\alpha}}{}^{\underline{\mu}}(x)\Lambda_{\underline{\beta}}{}^{\underline{\nu}}(x)\eta^{\underline{\alpha\beta}} = \eta^{\underline{\mu\nu}} \\ \Lambda_{\underline{\alpha}}{}^{\underline{\mu}}(x)\Lambda_{\underline{\beta}}{}^{\underline{\nu}}(x)\eta_{\underline{\mu\nu}} = \eta_{\underline{\alpha\beta}}\end{array}\right\}. \tag{51}$$

Now we performe the Lorentz transformation of the tetrad vectors $\{H^\mu_{\underline{\alpha}}(x)\}$ so that they coincided with $\{\tilde{H}^\mu_{\underline{\alpha}}(x)\}$, i.e. perform transformation of (50). At the transformation (50) the Dirac matrices $\gamma^\alpha(x)$ and $\gamma^{\underline{\alpha}}$ are transformed under the following rule:

$$\tilde{\gamma}^\alpha(x) = L(x)\gamma^\alpha(x)L^{-1}(x), \tag{52}$$

$$\gamma^{\underline{\alpha}} = \left[L(x)\gamma^{\underline{\beta}}L^{-1}(x)\right]\Lambda_{\underline{\beta}}{}^{\underline{\alpha}}(x) \tag{53}$$

The matrices $L$, $L^{-1}$ in (52), (53), are defined from the condition of invariance of the Dirac matrices $\gamma^{\underline{\alpha}}$ during the transformations (53), i.e. from the condition

$$L(x)\gamma^{\underline{\alpha}}L^{-1}(x) = \gamma^{\underline{\beta}}\Lambda_{\underline{\beta}}{}^{\underline{\alpha}}(x). \tag{54}$$

Since we are performing a Lorentz transformation, so that the system $\{H^\mu_{\underline{\alpha}}(x)\}$ coincides with the system $\{\tilde{H}^\mu_{\underline{\alpha}}(x)\}$, then the quantities $\Lambda_{\underline{\beta}}^{\underline{\alpha}}(x)$ in (54) should be assumed to equal to the corresponding quantities in (50). Note, that in our case the matrices $L$, $L^{-1}$ satisfy the relation

$$L(x)\gamma^0(x)L^{-1}(x) = \sqrt{-g^{00}}\,\gamma^{\underline{0}}, \tag{55}$$

which follows from the equalities (49) and (52).

The connection between the Hamiltonians (22) of the Schrödinger equation in an arbitrary gravitational field with the system of tetrad vectors $\{H^\mu_{\underline{\alpha}}(x)\}$ and $\{\tilde{H}^\mu_{\underline{\alpha}}(x)\}$ according to (49)-(53) is written in a standard form

$$\tilde{H} = LHL^{-1} + i\frac{\partial L}{\partial t}L^{-1}. \tag{56}$$

For stationary gravitational fields the Hamiltonian $H$ and the matrix $L$ do not depend on time, so the Hamiltonian in the Schwinger gauge is written as

$$\tilde{H} = LHL^{-1}. \tag{57}$$

Now we provide the explicit form of the matrix $L(x)$, satisfying the relation (55).

It is known that the Lorentz transformations can be unambiguously represented in the form of a product or boost transformation (Hermitian factor) by spatial rotation (unitary factor), either vice versa in the form of a product of the spatial rotation (unitary factor) by the boost transformation (Hermitian factor). This type is unambiguously factorized. Let us employ such factorization; we substitute (17) and (48) into (52) for $\tilde{\gamma}^0(x)$. In the result it turns out that the matrix $L(x)$ is written in the following form:

$$L(x) = R \cdot \exp\left\{\frac{\theta}{2}\cdot\frac{(\tilde{H}^\mu_{\underline{0}}H^\nu_{\underline{0}}S_{\mu\nu})}{\sqrt{(\tilde{H}^\varepsilon_{\underline{0}}H_{\underline{0}\varepsilon})^2 - 1}}\right\} = R\cdot\left\{\text{ch}\frac{\theta}{2} + \frac{(\tilde{H}^\mu_{\underline{0}}H^\nu_{\underline{0}}S_{\mu\nu})}{\sqrt{(\tilde{H}^\varepsilon_{\underline{0}}H_{\underline{0}\varepsilon})^2 - 1}}\cdot\text{sh}\frac{\theta}{2}\right\}. \tag{58}$$

Here $R$ represents the spatial rotation matrix, commutating with the matrix $\gamma^{\underline{0}}$. Another factor in (58) represents the hyperbolic rotation transformation (i.e. the boost) about the angle $\theta$, determined from the relation

$$\text{th}\frac{\theta}{2} = \sqrt{\frac{(\tilde{H}^\varepsilon_{\underline{0}}H_{\underline{0}\varepsilon}) + 1}{\sqrt{(\tilde{H}^\varepsilon_{\underline{0}}H_{\underline{0}\varepsilon}) - 1}}}. \tag{59}$$

The relation

$$\gamma^0(x) = L^{-1}(x)\tilde{\gamma}^0(x)L(x) \tag{60}$$

with account for (49) is written as

$$\gamma^0(x) = \left\{ \operatorname{ch}\frac{\theta}{2} - \frac{\left(\tilde{H}_{\underline{0}}^\mu H_{\underline{0}}^\nu S_{\mu\nu}\right)}{\sqrt{\left(\tilde{H}_{\underline{0}}^\varepsilon H_{\underline{0}\varepsilon}\right)^2 - 1}} \cdot \operatorname{sh}\frac{\theta}{2} \right\} \sqrt{-g^{00}}\, \gamma^{\underline{0}}(x) \exp\left\{ \frac{\theta}{2} \cdot \frac{\left(\tilde{H}_{\underline{0}}^\mu H_{\underline{0}}^\nu S_{\mu\nu}\right)}{\sqrt{\left(\tilde{H}_{\underline{0}}^\varepsilon H_{\underline{0}\varepsilon}\right)^2 - 1}} \right\}. \tag{61}$$

The relation (61) implies, that $L(x)$ is a matrix transforming $\gamma^0(x)$ into $\sqrt{-g^{00}}\,\gamma^{\underline{0}}$,

$$L\gamma^0 L^{-1} = \sqrt{-g^{00}}\,\gamma^{\underline{0}}. \tag{62}$$

### 5.3. Sense of the Operator $\eta$ at the use of the Parker Weight Operator

Substituting (27) into (62) with account for such properties of the matrices $L$ as

$$L^{-1} = -\gamma_{\underline{0}} L^+ \gamma_{\underline{0}}, \quad L^+ = -\gamma_{\underline{0}} L^{-1} \gamma_{\underline{0}}, \tag{63}$$

gives:

$$\rho = \sqrt{-g}\,\gamma_{\underline{0}}\gamma^0 = \sqrt{-g}\,\sqrt{-g^{00}}\, L^+ \cdot L. \tag{64}$$

The unitary matrix $R$, included into the matrix $L$ according to (58), is reduced in the product $L^+ \cdot L$ and does not affect the quantity $\rho$.

We can see that the operator $\rho$ can be written in the form (1), i.e. as

$$\rho = \eta^+ \eta. \tag{65}$$

At that the operator $\eta$ is proportional to the Lorentz transformation of $L$, which transforms $\gamma^0(x)$ into $\sqrt{-g^{00}}\gamma^{\underline{0}}$ according to (62),

$$\eta = (-g)^{1/4}\left(-g^{00}\right)^{1/4} \cdot L. \tag{66}$$

In case of using the system of tetrad vectors in the Schwinger gauge the operator $\eta$, defined by the relation (66), turns out to be equaling

$$\tilde{\eta} = (-g)^{1/4}\left(-g^{00}\right)^{1/4} \cdot E. \tag{67}$$

### 5.4. Uniqueness of the Hamiltonian $\mathrm{H}_\eta$ in the Case of Stationary Gravitational Fields

According to (4), (66) the Hamiltonian in the $\eta$ - representation can be written in the form

$$\mathrm{H}_\eta = \eta H \eta^{-1} = \tilde{\eta} L H L^{-1} \tilde{\eta}^{-1}. \tag{68}$$

Now let us chose the matrix $L(x)$ to have $\gamma^0(x) \to \sqrt{-g^{00}}\gamma^{\underline{0}}$ (see (60)-(62)). Then, according to (57) $LHL^{-1} = \tilde{H}$ and the expression (68) equals

$$\mathrm{H}_\eta = \tilde{\eta}\tilde{H}\tilde{\eta}^{-1} = \tilde{\mathrm{H}}_\eta. \tag{69}$$

For any system of tetrad vectors, after such operations will be obtained one and the same self-conjugate Hamiltonian (69) in the $\eta$-representation. It proves the uniqueness of the Hamil-tonian (69). This result confirms the results of the paper [1], in which for three systems of tetrad vectors after transition to the $\eta$-representation has been obtained one and the same self-conjugate Hamiltonian coincident with the Hamiltonian $\tilde{H}_\eta$ for the system of tetrad vectors in the Schwinger gauge.

## 6. SELF-CONJUGACY AND UNIQUENESS OF THE DIRAC HAMILTONIANS IN THE TIME-DEPENDENT GRAVITATIONAL FIELDS

Time-dependent gravitational fields do not satisfy the conditions of pseudo-Hermiticity of the Hamiltonians (2), (35).

However, the transition from the initial representation of the Dirac Hamiltonian $H$ to the $\eta$-representation allows obtaining the self-conjugate and unique Hamiltonian $H_\eta = H_\eta^+$ with the corresponding flat scalar product $(\Phi, \Psi)$, in the general case of the time-dependent gravitational field.

In fact, in the general case the Hamiltonian $H$ in the equation (5) depends on time

$$i\frac{\partial \psi}{\partial t} = H(t)\psi. \tag{70}$$

In this case the wave function $\Psi$ in the $\eta$-representation satisfies the equation

$$i\frac{\partial \Psi}{\partial t} = H_\eta \Psi = \left(\eta H \eta^{-1} + i\frac{\partial \eta}{\partial t}(\eta^{-1})\right)\Psi, \tag{71}$$

where we still have

$$\Psi = \eta \psi. \tag{72}$$

Now let us show the uniqueness of the Hamiltonian $H_\eta$ in the equation (71). Using (66), (67) gives

$$H_\eta = \eta H \eta^{-1} + i\frac{\partial \eta}{\partial t}\eta^{-1} = \\ = \tilde{\eta}\left(LHL^{-1} + i\frac{\partial L}{\partial t}L^{-1}\right)\tilde{\eta}^{-1} + i\frac{\partial \tilde{\eta}}{\partial t}\tilde{\eta}^{-1}. \tag{73}$$

We choose the transformation $L(x)$ so that $\gamma^0(x) \to \sqrt{-g^{00}}\gamma^{\underline{0}}$ (см. (60)-(62)). Then according to (56) we have

$$LHL^{-1} + i\frac{\partial L}{\partial t}L^{-1} = \tilde{H}, \tag{74}$$

$$H_\eta = \tilde{\eta}\tilde{H}\tilde{\eta}^{-1} + i\frac{\partial \tilde{\eta}}{\partial t}\tilde{\eta}^{-1}. \tag{75}$$

For any system of tetrad vectors, after the specified operations we will have one and the same Hamiltonian $H_\eta$ (75), what proves its uniqueness.

Now let us prove that the Hamiltonian $H_\eta$ in (74) is self-conjugate ($H_\eta = H_\eta^+$).

$$H_\eta^+ = \left( \tilde{\eta} \tilde{H} \tilde{\eta}^{-1} + i \frac{\partial \tilde{\eta}}{\partial t} \left( \tilde{\eta}^{-1} \right) \right)^+ = \left( \tilde{\eta}^{-1} \right)^+ \tilde{H}^+ \left( \tilde{\eta} \right)^+ - \left( \tilde{\eta}^{-1} \right)^+ i \frac{\partial \tilde{\eta}^+}{\partial t}. \quad (76)$$

For the system of tetrad vectors in the Schwinger gauge the relation (35) becomes equal

$$\tilde{H}^+ = \tilde{\rho} \tilde{H} \tilde{\rho}^{-1} + \tilde{\Delta};$$
$$\tilde{\rho} = (-g)^{1/2} (-g^{00})^{1/2}; \quad \tilde{\eta} = (-g)^{1/4} (-g^{00})^{1/4}; \quad \tilde{\gamma}^0 = \sqrt{-g^{00}} \gamma^0; \quad (77)$$
$$\tilde{\Delta} = i \frac{1}{2} \left( \frac{1}{(-g^{00})} \frac{\partial(-g^{00})}{\partial t} + \frac{1}{(-g)} \frac{\partial(-g)}{\partial t} \right) = 2i \frac{\partial \tilde{\eta}}{\partial t} \tilde{\eta}^{-1}.$$

Then the expression (76) equals

$$H_\eta^+ = \tilde{\eta} \tilde{H} \tilde{\eta}^{-1} + i \frac{\partial \tilde{\eta}}{\partial t} \left( \tilde{\eta}^{-1} \right) = H_\eta. \quad (78)$$

The scalar product in the $\eta$-representation is still flat and equals the initial one $\langle \varphi, \psi \rangle_\rho = (\Phi, \Psi)$.

## 7. ALGORITHM FOR FINDING THE HAMILTONIAN IN THE $\eta$-REPRESENTATION

Basing on the results of this work we can formulate the rules of finding the Hamiltonian in the $\eta$-representation for the Dirac particle in arbitrary gravitational field. The a priori information which we consider to be known is the information about the metric tensor $g_{\alpha\beta}(x)$, Christoffel symbols $\begin{Bmatrix} \lambda \\ \alpha\beta \end{Bmatrix}$, local metric tensor $\eta_{\underline{\alpha\beta}}$ and local Dirac matrices $\{ \gamma_{\underline{\alpha}} \}$. The specified rules consist in the following:

1) For the gravitational field described by the metric $g_{\alpha\beta}(x)$, it is necessary to find a system of tetrad vectors $\{ \tilde{H}_{\underline{\mu}}^\alpha(x) \}$, satisfying the Schwinger gauge. Note, that in this gauge the components of the tetrad vectors $\tilde{H}_{\underline{0}}^0, \tilde{H}_{\underline{0}}^k$ are connected with the components of the tensor $g^{\alpha\beta}(x)$ by the following relations:

$$\tilde{H}_{\underline{0}}^0 = \sqrt{-g^{00}}; \quad \tilde{H}_{\underline{0}}^k = -\frac{g^{0k}}{\sqrt{-g^{00}}}. \quad (79)$$

The components $\tilde{H}_{\underline{k}}^0$ are identically zero,

$$\tilde{H}_{\underline{k}}^0 = 0. \quad (80)$$

For finding $\tilde{H}_{\underline{m}}^n$ we introduce a tensor $f^{mn}$ with the components

$$f^{mn} = g^{mn} + \frac{g^{0m} g^{0n}}{g^{00}}. \quad (81)$$

The tensor $f^{mn}$ satisfies the conditions

$$f^{mn} g_{nk} = \delta_k^m. \quad (82)$$

Any three of three-dimensional vectors satisfying the relations written below will suit for the role of the quantities $\tilde{H}_{\underline{m}}^n$

$$\tilde{H}_{\underline{k}}^{m}\tilde{H}_{\underline{k}}^{n} = f^{mn}. \tag{83}$$

2) The general expression of the Hamiltonian $\tilde{H}$ is written:

$$\tilde{H} = -\frac{im}{\left(-g^{00}\right)}\tilde{\gamma}^{0} + \frac{i}{\left(-g^{00}\right)}\tilde{\gamma}^{0}\tilde{\gamma}^{k}\frac{\partial}{\partial x^{k}} - i\tilde{\Phi}_{0} + \frac{i}{\left(-g^{00}\right)}\tilde{\gamma}^{0}\tilde{\gamma}^{k}\tilde{\Phi}_{k}. \tag{84}$$

Here:

$$\tilde{\gamma}^{\alpha} = \tilde{H}_{\underline{\beta}}^{\alpha}\gamma^{\underline{\beta}}, \tag{85}$$

$$\tilde{\Phi}_{\alpha} = -\frac{1}{4}\tilde{H}_{\mu}^{\underline{\varepsilon}}\tilde{H}_{\nu\underline{\varepsilon};\alpha}S^{\mu\nu}. \tag{86}$$

3) According to (75)

$$\mathrm{H}_{\eta} = \tilde{\eta}\tilde{H}\tilde{\eta}^{-1} + i\tilde{\eta}\frac{\partial \tilde{\eta}^{-1}}{\partial t}, \tag{87}$$

where the operator $\tilde{\eta}$ is determined by the relation

$$\tilde{\eta} = \left(-g\right)^{1/4}\left(-g^{00}\right)^{1/4}. \tag{88}$$

The expressions (87), (88) define the operator $\mathrm{H}_{\eta}$, which is the searched Hermitian Hamiltonian in the $\eta$-representation.

So,

$$\begin{aligned}\mathrm{H}_{\eta} =& -\frac{im}{\left(-g^{00}\right)}\tilde{\gamma}^{0} + \frac{i}{\left(-g^{00}\right)}\tilde{\gamma}^{0}\tilde{\gamma}^{k}\frac{\partial}{\partial x^{k}} - i\tilde{\Phi}_{0} + \frac{i}{\left(-g^{00}\right)}\tilde{\gamma}^{0}\tilde{\gamma}^{k}\tilde{\Phi}_{k} \\ &- \frac{i}{4\left(-g^{00}\right)}\tilde{\gamma}^{0}\tilde{\gamma}^{k}\left\{\frac{\partial \ln(-g)}{\partial x^{k}} + \frac{\partial \ln(-g^{00})}{\partial x^{k}}\right\} + \frac{i}{4}\left\{\frac{\partial \ln(-g)}{\partial t} + \frac{\partial \ln(-g^{00})}{\partial t}\right\}.\end{aligned} \tag{89}$$

The rules specified above are applied further on for finding Hamiltonians in the $\eta$-representation for several stationary and non-stationary metrics.

## 8. Operators of the Hamiltonian in the $\eta$-Representation for Some Metrics

### 8.1. Metric used in the Works [3], [4]

Now let us consider the issue of constructing a Hamiltonian $\mathrm{H}_{\eta}$ of the stationary metric in the following form

$$ds^{2} = -V^{2}dt^{2} + W^{2}\left(dx^{2} + dy^{2} + dz^{2}\right), \tag{90}$$

where $V, W$ is the function of spatial coordinates. We will use the system of tetrad vectors in the Schwinger gauge:

$$\tilde{H}_{\underline{0}}^{0} = \frac{1}{V}, \quad \tilde{H}_{\underline{k}}^{0} = 0, \quad \tilde{H}_{\underline{0}}^{k} = 0, \quad \tilde{H}_{\underline{m}}^{k} = \frac{1}{W}\delta_{\underline{m}}^{k}. \tag{91}$$

The wave function $\psi$ evolution is determined according to the Schrodinger equation:

$$i\frac{\partial \psi}{\partial t} = \hat{H}\psi, \tag{92}$$

where $\hat{H}$ represents an operator of Hamiltonian (initial Hamiltonian). The explicit expression for the initial Hamiltonian $\hat{H}$ according to [3] has the form:

$$\tilde{H} = imV\gamma_{\underline{0}} - i\frac{V}{W}\gamma_{\underline{0}}\gamma^{\underline{k}}\frac{\partial}{\partial x^k} - \frac{i}{2}\cdot\frac{V_{,k}}{W}\cdot\gamma_{\underline{0}}\gamma^{\underline{k}} - i\frac{VW_{,k}}{W^2}\cdot\gamma_{\underline{0}}\gamma^{\underline{k}}. \tag{93}$$

In the considered case

$$\sqrt{-g} = VW^3; \quad \sqrt{-g^{00}} = \frac{1}{V}, \tag{94}$$

therefore

$$\tilde{\eta} = (-g)^{1/4}(-g^{00})^{1/4} = W^{3/2}. \tag{95}$$

In the $\eta$-representation the self-conjugate Hamiltonian $H_\eta$ is

$$H_\eta = \tilde{\eta}\tilde{H}\tilde{\eta}^{-1} = imV\gamma_{\underline{0}} - \frac{i}{2}\left(\gamma_{\underline{0}}\gamma^{\underline{k}}\frac{\partial}{\partial x^k}\frac{V}{W} + \frac{V}{W}\gamma_{\underline{0}}\gamma^{\underline{k}}\frac{\partial}{\partial x^k}\right). \tag{96}$$

The expression (96) was obtained without a supposition about the weakness of the gravitational field and it coincides with the self-conjugate Hamiltonian in papers [3].

## 8.2. SCHWARZSCHILD METRIC IN ISOTROPIC COORDINATES

The Schwarzschild metric is obtained from the metric (90) when writing the Schwarzschild solution in isotropic coordinates. In our discussion we omit the procedure of the corresponding coordinate transformation and show the results: for the transition of the function $V, W$ to isotropic coordinates a choice should be made according to the formulae (see, e.g., [3], [12]):

$$V = \frac{\left(1 - \frac{M}{2R}\right)}{\left(1 + \frac{M}{2R}\right)}; \quad W = \left(1 + \frac{M}{2R}\right)^2. \tag{97}$$

It follows from the expressions (96), (97) that the form of the Hamiltonian in the $\eta$-representation is:

$$H_\eta = im\frac{\left(1-\frac{M}{2R}\right)}{\left(1+\frac{M}{2R}\right)}\gamma_{\underline{0}} - i\frac{\left(1-\frac{M}{2R}\right)}{\left(1+\frac{M}{2R}\right)^3}\gamma_{\underline{0}}\gamma^{\underline{k}}\frac{\partial}{\partial x^k} - i\frac{\left(1-\frac{M}{4R}\right)}{\left(1+\frac{M}{2R}\right)^4}\frac{MR_{,k}}{R^3}\cdot\gamma_{\underline{0}}\gamma^{\underline{k}}. \tag{98}$$

In the case of weak fields the expression (98) in the lower order of approximation becomes equal to

$$H_\eta = im\gamma_{\underline{0}}\left(1 - \frac{M}{R}\right) - i\left(1 - 2\frac{M}{R}\right)\gamma_{\underline{0}}\gamma^{\underline{k}}\frac{\partial}{\partial x^k} - i\frac{MR_{,k}}{R^3}\cdot\gamma_{\underline{0}}\gamma^{\underline{k}}. \tag{99}$$

This expression coincides with that given in [1] (formula (58)). If necessary the exact expression (98) for the Hamiltonian can be expanded over the degrees of small parameter down to any infinitesimal order.

## 8.3. SCHWARZSCHILD METRIC IN THE COORDINATES $(t, r, \theta, \varphi)$

In this section we will write the Schwarzschild metric in the coordinates

$$\left(x^0, x^1, x^2, x^3\right) \equiv (t, r, \theta, \varphi). \tag{100}$$

In these coordinates we have:

$$ds^2 = -\left(1 - \frac{2M}{r}\right)dt^2 + \frac{dr^2}{\left(1 - \frac{2M}{r}\right)} + r^2\left(d\theta^2 + \sin^2\theta\, d\varphi^2\right). \tag{101}$$

For the metric determined by the quarter of the interval (101) we have:

$$g = -r^4 \cdot \sin^2\theta, \tag{102}$$

$$\left.\begin{aligned}
\tilde{\eta} &= (-g)^{1/4}\left(-g^{00}\right)^{1/4} = \frac{\left(r^4 \cdot \sin^2\theta\right)^{1/4}}{\left(1 - \frac{2M}{r}\right)^{1/4}}; \\
\tilde{\eta}\frac{\partial \tilde{\eta}^{-1}}{\partial r} &= -\frac{1}{r} + \frac{M}{2r^2} \cdot \frac{1}{\left(1 - \frac{2M}{r}\right)}; \quad \tilde{\eta}\frac{\partial \tilde{\eta}^{-1}}{\partial \theta} = -\frac{1}{2}\frac{\cos\theta}{\sin\theta}.
\end{aligned}\right\} \tag{103}$$

Now we define the tetrad vectors in the Schwinger gauge. Their nonzero components are

$$\left.\begin{aligned}
&\tilde{H}^0_{\underline{0}} = \frac{1}{\sqrt{f}}; \quad \tilde{H}^k_{\underline{0}} = 0; \quad \tilde{H}^0_{\underline{k}} = 0; \quad \tilde{H}^1_{\underline{1}} = \sqrt{f}; \quad \tilde{H}^2_{\underline{2}} = \frac{1}{r} \quad \tilde{H}^3_{\underline{3}} = \frac{1}{r \cdot \sin\theta}; \\
&\tilde{H}_{\underline{00}} = -\sqrt{f}; \quad \tilde{H}_{\underline{0k}} = 0; \quad \tilde{H}_{\underline{k0}} = 0; \quad \tilde{H}_{\underline{11}} = \frac{1}{\sqrt{f}}; \quad \tilde{H}_{\underline{22}} = r \quad \tilde{H}_{\underline{33}} = r \cdot \sin\theta; \\
&\tilde{H}^{\underline{00}} = -\frac{1}{\sqrt{f}}; \quad \tilde{H}^{\underline{0k}} = 0; \quad \tilde{H}^{\underline{k0}} = 0; \quad \tilde{H}^{\underline{11}} = \sqrt{f}; \quad \tilde{H}^{\underline{22}} = \frac{1}{r} \quad \tilde{H}^{\underline{33}} = \frac{1}{r \cdot \sin\theta}; \\
&\tilde{H}^{\underline{0}}_0 = \sqrt{f}; \quad \tilde{H}^{\underline{0}}_k = 0; \quad \tilde{H}^{\underline{k}}_0 = 0; \quad \tilde{H}^{\underline{1}}_1 = \frac{1}{\sqrt{f}}; \quad \tilde{H}^{\underline{2}}_2 = r \quad \tilde{H}^{\underline{3}}_3 = r \cdot \sin\theta.
\end{aligned}\right\} \tag{104}$$

Here $f \equiv 1 - \frac{2M}{r}$. The nonzero Christoffel symbols are:

$$\left.\begin{aligned}
&\binom{0}{01} = \frac{r_0}{2r^2\left(1 - \frac{r_0}{r}\right)}; \\
&\binom{1}{00} = \frac{r_0}{2r^2}\left(1 - \frac{r_0}{r}\right); \quad \binom{1}{11} = -\frac{r_0}{2r^2\left(1 - \frac{r_0}{r}\right)}; \quad \binom{1}{22} = -r\left(1 - \frac{r_0}{r}\right); \\
&\binom{1}{33} = -r \cdot \sin^2\theta \cdot \left(1 - \frac{r_0}{r}\right); \\
&\binom{2}{12} = \frac{1}{r}; \quad \binom{2}{33} = -\sin\theta\cos\theta; \quad \binom{3}{13} = \frac{1}{r}; \quad \binom{3}{23} = -\frac{\cos\theta}{\sin\theta}.
\end{aligned}\right\} \tag{105}$$

Using the formula (16) we define the quantities $\tilde{\Phi}_0$, $\tilde{\Phi}_k$.

$$\left.\begin{aligned}\tilde{\Phi}_0 &= \frac{M}{2r^2}\cdot\gamma_{\underline{0}}\gamma_{\underline{1}} \\ \tilde{\Phi}_1 &= 0 \\ \tilde{\Phi}_2 &= -\frac{1}{2}\sqrt{f}\cdot\gamma_{\underline{1}}\gamma_{\underline{2}} \\ \tilde{\Phi}_3 &= -\frac{1}{2}\cos\theta\cdot\gamma_{\underline{2}}\gamma_{\underline{3}}+\frac{1}{2}\sqrt{f}\sin\theta\cdot\gamma_{\underline{3}}\gamma_{\underline{1}}\end{aligned}\right\} \quad (106)$$

Now we substitute (104) into the expression (84) for $\tilde{H}$.

$$\begin{aligned}\tilde{H} &= im\sqrt{f}\gamma_{\underline{0}} - i\sqrt{f}\gamma_{\underline{0}}\left\{\gamma_{\underline{1}}\sqrt{f}\frac{\partial}{\partial r}+\gamma_{\underline{2}}\frac{1}{r}\frac{\partial}{\partial\theta}+\gamma_{\underline{3}}\frac{1}{r\cdot\sin\theta}\frac{\partial}{\partial\varphi}\right\} \\ &\quad -i\tilde{\Phi}_0 - i\sqrt{f}\gamma_{\underline{0}}\left\{\gamma_{\underline{1}}\sqrt{f}\tilde{\Phi}_1+\gamma_{\underline{2}}\frac{1}{r}\tilde{\Phi}_2+\gamma_{\underline{3}}\frac{1}{r\cdot\sin\theta}\tilde{\Phi}_3\right\}.\end{aligned} \quad (107)$$

After using the formulae (106) we have:

$$\begin{aligned}\tilde{H} &= im\sqrt{f}\gamma_{\underline{0}} - i\sqrt{f}\gamma_{\underline{0}}\left\{\gamma_{\underline{1}}\sqrt{f}\frac{\partial}{\partial r}+\gamma_{\underline{2}}\frac{1}{r}\frac{\partial}{\partial\theta}+\gamma_{\underline{3}}\frac{1}{r\cdot\sin\theta}\frac{\partial}{\partial\varphi}\right\} \\ &\quad -i\frac{M}{2r^2}\cdot\gamma_{\underline{0}}\gamma_{\underline{1}} - \frac{if}{r}\gamma_{\underline{0}}\gamma_{\underline{1}} - \frac{i\sqrt{f}\cos\theta}{2r\cdot\sin\theta}\gamma_{\underline{0}}\gamma_{\underline{2}}.\end{aligned} \quad (108)$$

We substitute (108) and (103) into the formula (87).

$$H_\eta = im\sqrt{f}\gamma_{\underline{0}} - i\sqrt{f}\gamma_{\underline{0}}\left\{\gamma_{\underline{1}}\sqrt{f}\frac{\partial}{\partial r}+\gamma_{\underline{2}}\frac{1}{r}\frac{\partial}{\partial\theta}+\gamma_{\underline{3}}\frac{1}{r\cdot\sin\theta}\frac{\partial}{\partial\varphi}\right\} - \frac{i}{2}\frac{\partial f}{\partial r}\cdot\gamma_{\underline{0}}\gamma_{\underline{1}}. \quad (109)$$

The expression (109) is an operator of the Hamiltonian in the $\eta$-representation. It is easy to see that this operator is Hermitian ($H_\eta = H_\eta^+$).

Note, that the expressions (109) and (98) are equivalent to each other, only one and the same Hamiltonian is written in different reference frames.

### 8.4. Friedmann Models

Now let us consider the Schrödinger equation in the case when the space is homogeneous and isotropic, and hence, it is described by some solution of the Friedmann model. We restrict ourselves by two elementary solutions – the cosmologically flat and the open Friedmann models [11].

#### Cosmologically Flat Friedmann Model

The nonstationary metric which corresponds to the cosmologically flat Friedmann solution is determined by the relation:

$$ds^2 = -dt^2 + b^2(t)\left[dx^2 + dy^2 + dz^2\right]. \quad (110)$$

The Christoffel symbols corresponding to (110), are:

$$\left.\begin{aligned}\binom{0}{00} &= 0;\quad \binom{0}{0k}=0;\quad \binom{0}{mn}=b\dot{b}g_{mn}; \\ \binom{k}{00} &= 0;\quad \binom{m}{0n}=\frac{\dot{b}}{b}\delta_n^m;\quad \binom{k}{mn}=0.\end{aligned}\right\} \quad (111)$$

The tetrad vectors $\tilde{H}^\alpha_\alpha$ in the Schwinger gauge are determined according to their relation (11) and by using the expression (110).

$$\left.\begin{array}{l}\tilde{H}^{\underline{0}}_\alpha=(1,0,0,0);\ \tilde{H}^{\underline{1}}_\alpha=(0,b,0,0);\ \tilde{H}^{\underline{2}}_\alpha=(0,0,b,0);\ \tilde{H}^{\underline{3}}_\alpha=(0,0,0,b);\\ \tilde{H}_{\underline{0}\alpha}=(-1,0,0,0);\ \tilde{H}_{\underline{1}\alpha}=(0,b,0,0);\ \tilde{H}_{\underline{2}\alpha}=(0,0,b,0);\ \tilde{H}_{\underline{3}\alpha}=(0,0,0,b);\\ \tilde{H}^{\underline{0}\alpha}=(-1,0,0,0);\ \tilde{H}^{\underline{1}\alpha}=\left(0,\frac{1}{b},0,0\right);\ \tilde{H}^{\underline{2}\alpha}=\left(0,0,\frac{1}{b},0\right);\ \tilde{H}^{\underline{3}\alpha}=\left(0,0,0,\frac{1}{b}\right);\\ \tilde{H}^\alpha_{\underline{0}}=(1,0,0,0);\ \tilde{H}^\alpha_{\underline{1}}=\left(0,\frac{1}{b},0,0\right);\ \tilde{H}^\alpha_{\underline{2}}=\left(0,0,\frac{1}{b},0\right);\ \tilde{H}^\alpha_{\underline{3}}=\left(0,0,0,\frac{1}{b}\right).\end{array}\right\} \quad (112)$$

Calculating the component $\Phi_\alpha$ using (110)-(112) gives:

$$\left.\begin{array}{l}\tilde{\Phi}_0=0\\ \tilde{\Phi}_k=\frac{1}{2}\frac{\dot{b}}{b}\cdot S_{0k}\end{array}\right\}. \quad (113)$$

Now we substitute the expressions (113) for the components $\tilde{\Phi}_\alpha$ into the Hamiltonian and get:

$$\tilde{H}=im\tilde{\gamma}_{\underline{0}}-i\tilde{\gamma}_{\underline{0}}\tilde{\gamma}^k\frac{\partial}{\partial x^k}-\frac{3i}{2}\frac{\dot{b}}{b}. \quad (114)$$

Since in this case

$$\tilde{\gamma}_{\underline{0}}=\gamma_{\underline{0}};\quad \tilde{\gamma}^k=\frac{1}{b}\gamma^{\underline{k}}=\frac{1}{b}\gamma_{\underline{k}}, \quad (115)$$

The Hamiltonian (114) can be written in the following form

$$\tilde{H}=im\gamma_{\underline{0}}-\frac{i}{b}\gamma_{\underline{0}}\gamma_{\underline{k}}\frac{\partial}{\partial x^k}-\frac{3i}{2}\frac{\dot{b}}{b}. \quad (116)$$

Then we will get the Hamiltonian of the considered system in the $\eta$-representation. The Parker weight operation and the operator $\tilde{\eta}$ are:

$$\tilde{\rho}=\sqrt{-g}\gamma_{\underline{0}}\tilde{\gamma}^0=b^3;\ \tilde{\eta}=b^{3/2}. \quad (117)$$

Then

$$H_\eta=\tilde{\eta}\tilde{H}\tilde{\eta}^{-1}+i\frac{\partial\tilde{\eta}}{\partial t}\left(\tilde{\eta}^{-1}\right)=im\gamma_{\underline{0}}-\frac{i}{b(t)}\gamma_{\underline{0}}\gamma_{\underline{k}}\frac{\partial}{\partial x^k}=H^+_\eta. \quad (118)$$

According to the result of Section 6 after the transition to the $\eta$-representation the initially non-Hermitian Hamiltonian (114) is transformed into a self-conjugate Hamiltonian (118) with the corresponding flat scalar product.

The operator of the Dirac particle energy in the $\eta$-representation is

$$E=\sqrt{H_\eta^2}=\sqrt{m^2+\frac{\mathbf{p}^2}{b^2(t)}}. \quad (119)$$

In the expression (119) $p^k=-i\frac{\partial}{\partial x^k}$ represents the components of the Dirac particle momentum.

## OPEN FRIEDMANN MODEL

Let us consider the case of the open Friedmann model in the coordinates:
$$\left(x^0,x^1,x^2,x^3\right)=(t,\chi,\theta,\varphi).$$
For this model the nonstationary metric has the form:

$$ds^2=-dt^2+a^2(t)\left(d\chi^2+\operatorname{sh}^2\chi\left[d\theta^2+\sin^2\theta d\varphi^2\right]\right). \quad (120)$$

The nonzero Christoffel symbols corresponding to the metric (120) have the form:

$$\begin{pmatrix} 0 \\ 00 \end{pmatrix} = 0; \quad \begin{pmatrix} 0 \\ 0k \end{pmatrix} = 0; \quad \begin{pmatrix} 0 \\ mn \end{pmatrix} = a\dot{a}g_{mn};$$

$$\begin{pmatrix} k \\ 00 \end{pmatrix} = 0; \quad \begin{pmatrix} m \\ 0n \end{pmatrix} = \frac{\dot{a}}{a}\delta_n^m; \quad \begin{pmatrix} 1 \\ 22 \end{pmatrix} = -\text{sh}\chi \cdot \text{ch}\chi;$$

$$\begin{pmatrix} 1 \\ 33 \end{pmatrix} = -\text{sh}\chi \cdot \text{ch}\chi \cdot \sin^2\theta; \quad \begin{pmatrix} 2 \\ 12 \end{pmatrix} = \frac{\text{ch}\chi}{\text{sh}\chi}; \quad \begin{pmatrix} 3 \\ 13 \end{pmatrix} = \frac{\text{ch}\chi}{\text{sh}\chi};$$

$$\begin{pmatrix} 2 \\ 33 \end{pmatrix} = -\sin\theta\cos\theta; \quad \begin{pmatrix} 3 \\ 23 \end{pmatrix} = \text{ctg}\,\theta. \quad \quad (121)$$

The nonzero components of the tetrad vectors $\tilde{H}_{\underline{\alpha}}^{\alpha}$ in the Schwinger gauge are:

$$\begin{cases} \tilde{H}_{\underline{0}}^0 = 1; \tilde{H}_{\underline{1}}^1 = \frac{1}{a}; \tilde{H}_{\underline{2}}^2 = \frac{1}{a \cdot \text{sh}\chi}; \tilde{H}_{\underline{3}}^3 = \frac{1}{a \cdot \text{sh}\chi \cdot \sin\theta}; \\ \tilde{H}_{\underline{00}} = -1; \tilde{H}_{\underline{11}} = a; \tilde{H}_{\underline{22}} = a \cdot \text{sh}\chi; \tilde{H}_{\underline{33}} = a \cdot \text{sh}\chi \cdot \sin\theta; \\ \tilde{H}_{\underline{\alpha}}^0 = 1; \tilde{H}_{\underline{1}}^1 = a; \tilde{H}_{\underline{2}}^2 = a \cdot \text{sh}\chi; \tilde{H}_{\underline{3}}^3 = a \cdot \text{sh}\chi \cdot \sin\theta; \\ \tilde{H}^{\underline{00}} = -1; \tilde{H}^{\underline{11}} = \frac{1}{a}; \tilde{H}^{\underline{22}} = \frac{1}{a \cdot \text{sh}\chi}; \tilde{H}^{\underline{33}} = \frac{1}{a \cdot \text{sh}\chi \cdot \sin\theta}. \end{cases} \quad (122)$$

The operator $\tilde{\eta}$ is

$$\tilde{\eta} = (-g)^{1/4} \left(-g^{00}\right)^{1/4} = \left(a^6 \text{sh}^4\chi \sin^2\theta\right)^{1/4}. \quad (123)$$

The quantities $\tilde{\eta}\frac{\partial \tilde{\eta}^{-1}}{\partial x^k}$, required for finding the Hamiltonian in the $\eta$-representation are:

$$\tilde{\eta}\frac{\partial \tilde{\eta}^{-1}}{\partial \chi} = -\text{cth}\,\chi; \quad \tilde{\eta}\frac{\partial \tilde{\eta}^{-1}}{\partial \theta} = -\frac{1}{2}\text{ctg}\,\theta; \quad \tilde{\eta}\frac{\partial \tilde{\eta}^{-1}}{\partial \varphi} = 0. \quad (124)$$

Calculation of the components of $\tilde{\Phi}_\alpha$ using the expression (122) and (121) shows that

$$\begin{aligned} \tilde{\Phi}_0 &= 0, \\ \tilde{\Phi}_1 &= \frac{\dot{a}}{2} \cdot \gamma_{\underline{0}}\gamma_{\underline{1}}, \\ \tilde{\Phi}_2 &= \frac{\dot{a}}{2}\text{sh}\chi \cdot \gamma_{\underline{0}}\gamma_{\underline{2}} - \frac{1}{2}\text{ch}\chi \cdot S^{\underline{12}}, \\ \tilde{\Phi}_3 &= \frac{\dot{a}}{2}\text{sh}\chi\sin\theta \cdot \gamma_{\underline{0}}\gamma_{\underline{3}} + \frac{1}{2}\text{ch}\chi \cdot \sin\theta \cdot S^{\underline{31}} - \frac{1}{2}\cos\theta \cdot S^{\underline{23}}. \end{aligned} \quad (125)$$

Calculation of the Hamiltonian $\tilde{H}$ gives:

$$\tilde{H} = im\gamma_{\underline{0}} - i\gamma_{\underline{0}}\gamma_{\underline{1}}\frac{1}{a}\frac{\partial}{\partial\chi} - i\gamma_{\underline{0}}\gamma_{\underline{2}}\frac{1}{a \cdot \text{sh}\chi}\frac{\partial}{\partial\theta} - i\gamma_{\underline{0}}\gamma_{\underline{3}}\frac{1}{a \cdot \text{sh}\chi \cdot \sin\theta}\frac{\partial}{\partial\varphi}$$
$$-\frac{i}{a}\text{cth}\chi \cdot \gamma_{\underline{0}}\gamma_{\underline{1}} - \frac{i}{2a}\frac{\text{ctg}\,\theta}{\text{sh}\chi}\cdot\gamma_{\underline{0}}\gamma_{\underline{2}} - i\frac{3}{2}\frac{\dot{a}}{a}. \quad (126)$$

Now we calculate the operator $H_\eta$, using (89), (123) and (124). We have:

$$H_\eta = im\gamma_{\underline{0}} - i\gamma_{\underline{0}}\gamma_{\underline{1}}\frac{1}{a}\frac{\partial}{\partial\chi} - i\gamma_{\underline{0}}\gamma_{\underline{2}}\frac{1}{a \cdot \text{sh}\chi}\frac{\partial}{\partial\theta} - i\gamma_{\underline{0}}\gamma_{\underline{3}}\frac{1}{a \cdot \text{sh}\chi \cdot \sin\theta}\frac{\partial}{\partial\varphi}. \quad (127)$$

The quantity $H_\eta$, determined by the relation (127), is a Hamiltonian in the $\eta$-representation for the Dirac particles in the open Friedmann model. After transition to the $\eta$-representation the initially non-Hermitian Hamiltonian (126) is transformed into a self-conjugate Hamiltonian (127) with the corresponding flat scalar product.

The operator of the energy for a particle moving in the $\chi$-direction is:

$$E = \sqrt{H_\eta^2} = \sqrt{m^2 + \frac{\mathbf{p}_\chi^2}{a^2(t)}}. \qquad (128)$$

Here $\mathbf{p}_\chi = -i\dfrac{\partial}{\partial \chi}$.

Denote

$$a(t)\operatorname{sh}\chi = \frac{a(t)}{a_0} a_0 \operatorname{sh}\chi = b(t) a_0 \operatorname{sh}\chi = b(t) r, \qquad (129)$$

where $b(t_0) = 1$. The zero indices correspond to the present time $(t \leq t_0)$.

If at the present time the radius of the Universe spatial curvature tends to infinity $(a_0 \to \infty)$, then

$$r \approx a_0 \chi. \qquad (130)$$

In this case the Hamiltonian (127) becomes equal to

$$\begin{aligned}H_\eta &= im\gamma_{\underline{0}} - i\gamma_{\underline{0}}\gamma_{\underline{1}} \frac{1}{b(t)} \frac{\partial}{\partial r} - i\gamma_{\underline{0}}\gamma_{\underline{2}} \frac{1}{b(t)r} \frac{\partial}{\partial \theta} - i\gamma_{\underline{0}}\gamma_{\underline{3}} \frac{1}{b(t)r\sin\theta} \frac{\partial}{\partial \varphi} = \\ &= im\gamma_{\underline{0}} - i\gamma_{\underline{0}}\gamma_{\underline{k}} \frac{1}{b(t)} (\nabla_k)_{sph}.\end{aligned} \qquad (131)$$

In the expression (131) the quantity $(\nabla_k)_{sph}$ represents the gradient components in the spherical reference frame. Apparently, in the Cartesian frame of reference the Hamiltonian (131) coincides with the Hamiltonian (118) for cosmologically flat Friedmann model.

The physical implication of the Hamiltonians (118), (131) for the Dirac particles in the expanding Universe will be presented in the next work of the authors. The major results are as follows:

1) The Hamiltonians (118), (131) do not result in additional cosmological shift of the atomic spectral lines when the interaction with electro-magnetic field is taken into account. It is consistent with the modern cosmological model, $\Lambda CDM$ ("concordance model").

2) The Universe expansion results in the cosmological change of the interaction forces of elementary particles.

## 9. CONCLUSIONS

The results of the present work allow us to draw a conclusion that the problem of uniqueness and self-conjugacy of the Dirac Hamiltonians in arbitrary gravitational fields, both stationary and time dependent, is solved.

The unique properties of the Parker weight operator $\rho = \sqrt{-g}\gamma_{\underline{0}}\gamma^0 = \eta^+\eta$ allow obtaining in the $\eta$-reprsentation uniquely self-conjugate Hamiltonians of Dirac particles in arbitrary gravitational fields.

This conclusion is true both for the case of fulfillment the pseudo-Hermiticity condition (2), when the initial Hamiltonian is Hermitian with respect to the Parker scalar product (stationary gravitational fields) and in the case of violation of the condition (2), when the initial Hamiltonian is non-Hermitian with respect to the Parker scalar product (nonstationary gravitational fields). In the latter case the transition of the system of tetrad vectors in the Schwinger gauge is required for obtaining self-conjugate Hamiltonians of Dirac particles.

The scalar products in the $\eta$-representation are flat which allows using conventional apparatus of the Hermitian quantum mechanics. Evidently the observed physical quantities in the initial representations should be properly transformed at the transition to the $\eta$-representation $\left(O \rightarrow \eta O \eta^{-1}\right)$.

Basing on the presented discussion the rules (the general algorithm) for finding the Hamiltonian in the $\eta$-representation, which are applicable for any kind of gravitational field, are formulated in Section 7. The general approach is demonstrated at the deriving the equations for the Dirac Hamiltonian for stationary metric considered in paper [3], stationary Schwarzschild solution in isotropic coordinates and in the coordinates $(t,r,\theta,\varphi)$, as well as for nonstationary cosmologically flat and the open Friedmann models.

# REFERENCES

[1] M. V. Gorbatenko, V. P. Neznamov. Phys. Rev. D82, 104056 (2010); arXiv: 1007.4631v1[gr-qc].

[2] F. W. Hehl and W. T. Ni. Phys. Rev. D42. 2045 (1990).

[3] Yu. N. Obukhov. Phys. Rev. Lett. **86**, 192 (2001); Forschr. Phys. **50**, 711 (2002); arXiv: gr-qc/0012102.

[4] Yu. N. Obukhov, A. J. Silenko, O. V. Teryaev. Phys. Rev. D**80**, 064044 (2009), arXiv: 0907.4367v1[gr-qc].

[5] L. Parker. Phys. Rev. D22. 1922 (1980).

[6] Xing Huang, L. Parker. Phys. Rev. **D**79, 024020 (2009); arXiv: 0811.2296v1[hep-th].

[7] C. M. Bender, D. Brody and H. F. Jones. Phys.Rev.Lett. **89,** 2704041 (2002); Phys. Rev. D70, 025001 (2004).

[8] A. Mostafazadeh. J.Math Phys. (N.Y.) **43**, 205 (2002), **43,** 2814 (2002), **43,** 3944 (2002); arXiv: 0810.5643v3[quant-ph].

[9] B. Bagchi, A. Fring. Phys. Lett. A 373, 4307 (2009), arXiv: hep-th/0907.5354v1.

[10]     J. Schwinger. Phys. Rev. 130 (1963) 800-805.

[11]     L. D. Landau and E. M. Lifshitz. *The Classical Theory of Fields*. Pergamon Press, Oxford (1975).

[12]     A. Lihgtman, W. Press, R. Price, S. Teukolsky. *Problem Book in Relativity and Gravitation* (Princeton University Press, Princeton, New Jersey, 1975).